%% file: 0_main.tex
\documentclass[sigconf, authorversion]{acmart}

\AtBeginDocument{%
  \providecommand\BibTeX{{%
    \normalfont B\kern-0.5em{\scshape i\kern-0.25em b}\kern-0.8em\TeX}}}

\begin{document}

\title{Assessing Algorithmic Biases for Musical Version Identification}

\author{Furkan Yesiler}
\email{furkan.yesiler@upf.edu}
\affiliation{%
  \institution{Music Technology Group, Pompeu Fabra University}
  \city{Barcelona}
  \country{Spain}
}

\author{Marius Miron}
\affiliation{%
  \institution{Music Technology Group, Pompeu Fabra University}
  \city{Barcelona}
  \country{Spain}
}

\author{Joan Serr\`a}
\affiliation{%
  \institution{Dolby Laboratories}
  \city{Barcelona}
  \country{Spain}
}

\author{Emilia G\'omez}
\affiliation{%
  \institution{Joint Research Centre, European Commission}
  \city{Sevilla}
  \country{Spain}
}

\renewcommand{\shortauthors}{Yesiler, et al.}

\begin{abstract}
Version identification (VI) systems now offer accurate and scalable solutions for detecting different renditions of a musical composition, allowing the use of these systems in industrial applications and throughout the wider music ecosystem.
Such use can have an important impact on various stakeholders regarding recognition and financial benefits, including how royalties are circulated for digital rights management.
In this work, we take a step toward acknowledging this impact and consider VI systems as socio-technical systems rather than isolated technologies.
We propose a framework for quantifying performance disparities across 5~systems and 6~relevant side attributes: gender, popularity, country, language, year, and prevalence.
We also consider 3~main stakeholders for this particular information retrieval use case: the performing artists of query tracks, those of reference (original) tracks, and the composers.
By categorizing the recordings in our dataset using such attributes and stakeholders, we analyze whether the considered VI systems show any implicit biases.
We find signs of disparities in identification performance for most of the groups we include in our analyses. 
Moreover, we also find that learning- and rule-based systems behave differently for some attributes, which suggests an additional dimension to consider along with accuracy and scalability when evaluating VI systems.
Lastly, we share our dataset with attribute annotations to encourage VI researchers to take these aspects into account while building new systems.

\end{abstract}



\newcommand\datasetname{VI-Bias}
\newcommand\rrsym{$\psi$}
\newcommand\mrrsym{$\bar{\psi}$}
\newcommand\fy[1]{\textbf{\textcolor{cyan}{#1}}}

\newcommand\glpair[2]{{\small\textsf{#1}}-{\scriptsize\textsf{#2}}}
\newcommand\sigone{\redt{*}}
\newcommand\sigtwo{\redt{**}}
\newcommand{\cmark}{\textcolor{green}{\ding{51}}}%
\newcommand{\xmark}{\textcolor{red}{\ding{53}}}
\newcommand\redt[1]{\textbf{\textcolor{red}{#1}}}
\newcommand\bluet[1]{#1}
\maketitle

\input{1_introduction}
\input{2_methodology}
\input{3_results}
\input{4_discussion}
\input{5_conclusion}

\begin{acks}
This work is supported by the MIP-Frontiers project, the European Union's Horizon 2020 research and innovation programme under the Marie Skłodowska-Curie grant agreement No.~765068.
\end{acks}

\bibliographystyle{ACM-Reference-Format}
\bibliography{sample-base}

\input{6_appendix}

\end{document}

%% file: 1_introduction.tex
\section{Introduction}\label{sec:introduction}

Version identification (VI) refers to the task of automatically detecting different audio renditions of a musical composition (e.g.,~cover songs). Considering that versions can change in several musical characteristics (including key, tempo, structure, and recording conditions), VI systems aim to look past such changes and focus on the information that is common between versions~\cite{serra2010audio}. 
With improved accuracy and scalability, VI systems start being used in commercial applications, ranging from music recognition to music discovery, and such usage can have an impact on various stakeholders in the music ecosystem, especially regarding recognition and financial benefits.

Digital rights management (DRM) for musical recordings is a complex ecosystem, where multiple types of royalties circulate between a large amount of stakeholders~\cite{royalties}. 
When artists perform a version of an existing composition, they are obliged to pay ``mechanical royalties'' to the holders of ``the composition copyright.'' 
With the help of VI systems, detecting such cases could be automated. 
However, VI systems cannot match a recording to a composition directly; instead, they match an unknown recording to the known recordings of a composition. 
Therefore, we can identify three main parties involved in the process: the performing artists of the unknown (or query) recording, the performing artists of the known (or reference) recording, and the composers of the composition.
Taking into account the impact that VI systems may have over such artists and composers, they can be considered as socio-technical systems~\cite{baxter2011, selbst2019fairness} rather than isolated technologies for such use cases.

Fairness and transparency is an emerging field that studies the societal implications of algorithmic systems~\cite{olteanu2021facts, mehrabi2021survey}.
Any existing biases in such systems may lead to favor certain individuals or groups over others, which may result in ``unfair'' outcomes~\cite{fletcher2021}. 
Although algorithmic decisions and their impact are closely related concepts, measuring fairness is a domain- and context-dependent process, mainly due to the fact that investigating societal impact requires a certain legal, cultural, and ethical framework. However, quantifying algorithmic biases is a useful step for mitigating potential unfair decisions.
In music technology research, recent works have addressed potential issues in music recommendation from both individual and group fairness perspectives~\cite{chouldechova2018frontiers, raz2021} by studying gender imbalance~\cite{ferraro2021break,shakespeare2020exploring} and playlist diversity~\cite{porcaro201920,robinson2020user}.
In a recommendation context, a fairness study may investigate whether two potentially impacted groups are subject to the same exposure or not, assuming more exposure leads to larger financial gains.
Nonetheless, in a music recognition task like VI, measuring fairness with respect to such exposure may not be meaningful, as the main goal of VI systems is to correctly detect items that are objectively linked to each other as ``versions of the same composition.''
However, this does not imply that tasks like VI, which have objectively assigned labels, do not require any studies about fairness or algorithmic biases.
For instance, a discrepancy in the identification performance of such VI systems with respect to some characteristics (e.g.,~demographics) of the involved parties (e.g.,~performing artists or composers) may put some musicians in a favored position (e.g.,~financially) compared to others. 
Therefore, an examination into whether VI systems inherit any form of bias is necessary to study the potential implications of such systems in the music ecosystem.

In this work, we propose a framework for investigating the performance discrepancies of VI systems across a selected set of attributes, in order to explore their impact on relevant parties.
Although our main motivation is to raise awareness for developing fair VI systems, we cannot reach any conclusions on the fairness qualities of existing systems since such fairness measurements are context-dependent and we do not have access to a system that is interacting with the parties we consider.
Instead, we aim to quantify the algorithmic bias by investigating the discrepancies in identification performances of 5~state-of-the-art VI systems of different characteristics on pairs of potentially impacted groups, mimicking a group fairness paradigm~\cite{raz2021}.
We categorize such groups using the 3~aforementioned parties (performing artists and composers) and 6~relevant side attributes (gender, popularity, country, language, year, and prevalence).
To carry out our analyses, we develop the \datasetname\ dataset, for which we collect metadata in terms of the attributes we investigate and assign them to songs of an existing dataset.
We then gather performance data on pairs of groups and analyze the existence of potential discrepancies using the Kolmogorov-Smirnov test.
Our results from a total of 115~experiments suggest that, on the one hand, the characteristics of a VI system may play a role in favoring a certain group over its counterpart but that, on the other hand, there also exist cases where all the VI systems favor the same or no group.
After presenting the results, we also share our hypotheses on the possible reasons for the observed disparities.
To facilitate future research, we share the instructions on how to obtain our dataset and the evaluation code\footnote{\url{https://github.com/furkanyesiler/vi_bias}}.

%% file: 2_methodology.tex
\section{Methodology}\label{sec:methodology}

\subsection{Systems}\label{ssec:systems}

In this study, we consider a range of VI systems with different characteristics, in terms of whether they are learning- or rule-based, or whether they use melody- or chroma-based inputs. With this, we aim to better understand and discuss the causes of possible performance differences.

\subsubsection{Qmax} 
Qmax~\cite{serra2009cross} is a rule-based system that estimates similarities between pairs of tracks using an elaborate local alignment scheme, which results in a good performance but suffers from high computational cost. Nevertheless, it was the state-of-the-art system for VI for more than a decade, until recently. For our experiments, we use the implementation included in Essentia~\cite{essentia}, with cremaPCP features~\cite{yesiler2019tacos} as input, which are specialized chroma representations that correspond to an intermediate output of a structured chord estimation model~\cite{mcfee2017structured}.

\subsubsection{MOVE} MOVE~\cite{yesiler2019accurate} is a deep learning-based system that features musically motivated design decisions to bring inductive biases to the model. It also uses cremaPCP features as input and transforms them into fixed-size embedding vectors, regardless of the input duration. By having explicit modules for achieving transposition and structure invariance, it has been proven effective for VI.

\subsubsection{MICE-F and MICE-C} MICE~\cite{doras2020combining} is a recent deep learning-based system that combines the design decisions of the model introduced by Doras et al.~\cite{doras2019cover} and MOVE. Due to its input-agnostic design, MICE can be trained and used with any input feature suitable for VI. Hence, we consider two variants of the algorithm: MICE-F, which uses dominant melody features as input, and MICE-C, which uses cremaPCP features.

\subsubsection{LF-c} LF-c~\cite{doras2020combining} is a deep learning-based system designed to investigate the performance gains of combining models that use complementary input features (i.e.,~dominant melody and cremaPCP). It concatenates the embeddings obtained by MOVE and MICE-F and projects these into a new space, resulting in new embeddings. Doras et al.~\cite{doras2020combining} show that using such an ensemble system improves identification performance by a large margin. 

\subsection{Attributes}\label{ssec:attributes}
For each of the attributes presented below, we categorize recordings into two groups. Considering various parties involved in VI, these groups can be created with respect to (1)~the artist or the performance of the query track, (2)~the artist or the performance of the reference or ``original'' track, and (3)~the composer or the composition. 

In the case of attributes that relate to persons (e.g.,~gender) rather than recordings (e.g.,~language), if the artist of a recording is a person, we assign the label for that recording as that of that person. If the artist is a band, we then collect the labels for all the current and past band members. To have clear distinctions between groups, we do not consider bands that have mixed labels (i.e.,~where some members belong to one group while the others belong to the other group). We also exclude from our analyses the cases where we cannot find a label for even a single member of a band.

We study differences in system performances by using binary labels (group 1, or G1, and group 2, or G2) for each attribute. If not stated otherwise, such labels are derived with respect to (1)~the query recordings or their artists (\glpair{}{Q}), (2)~the reference recordings or their artists (\glpair{}{R}), and (3)~the compositions or their composers (\glpair{}{C}). We also analyze the performance differences where the query and the reference recordings or their artists belong to the same vs.\ different groups (\glpair{}{SD}). In the case where they belong to different groups, we also check if the direction of the change (i.e.,~the reference in G1 and the query in G2 or the reference in G2 and the query in G1) has any effect on the performance (\glpair{}{D12}). A summary of all the experiments and the considered attributes can be seen in Table~\ref{tab:exp-summary}. 

Note that, for the \glpair{}{SD} experiments, G1 denotes the cases where queries and references belong to the same group (e.g.,~queries and references from male artists) while G2 denotes that they belong to different groups (e.g.,~queries from male artists and references from female artists). For the \glpair{}{D12} experiments, G1 denotes the cases where queries and references are in the first and the second groups of the \glpair{}{Q} experiments, respectively (e.g.,~queries from male artists and references from female artists); and G2 denotes cases where the queries and references are in the second and the first groups of the \glpair{}{Q} experiments, respectively (e.g.,~queries from female artists and references from male artists). 

\begin{table}[tb!]
\caption{Summary of the experiments. Numbers indicate the considered systems: Qmax (1), MOVE (2), MICE-F (3), MICE-C (4), and LF-c (5). The color \redt{red} indicates that the results obtained with the corresponding system show a significant difference between G1 and G2.
}\label{tab:exp-summary}
\begin{center}
\resizebox{\columnwidth}{!}{
\begin{tabular}{l c c c c c}
\toprule
Feature & Query & Ref. & Comp. & Same-Dif. & Dif.\,G \\
 & (\glpair{}{Q}) & (\glpair{}{R}) & (\glpair{}{C}) & (\glpair{}{SD}) & (\glpair{}{D12}) \\
\midrule

Gender (\glpair{G}{}) & \bluet{1}\redt{2}\redt{3}\redt{4}\redt{5} & 
\bluet{1}\redt{2}\redt{3}\redt{4}\redt{5} & 
\bluet{1}\bluet{2}\redt{3}\bluet{4}\redt{5} & 
\bluet{1}\redt{2}\redt{3}\redt{4}\redt{5} & 
\bluet{1}\bluet{2}\bluet{3}\bluet{4}\bluet{5} 
\\

Popularity (\glpair{P}{}) & \bluet{1}\redt{2}\redt{3}\redt{4}\redt{5} & 
\redt{1}\redt{2}\redt{3}\redt{4}\redt{5} & 
\bluet{1}\bluet{2}\bluet{3}\bluet{4}\bluet{5} & 
\redt{1}\redt{2}\redt{3}\redt{4}\redt{5} & 
\bluet{1}\bluet{2}\bluet{3}\bluet{4}\bluet{5} 
\\

Country (\glpair{C}{})  & \bluet{1}\bluet{2}\bluet{3}\bluet{4}\bluet{5} & 
\bluet{1}\bluet{2}\bluet{3}\bluet{4}\bluet{5} & 
\bluet{1}\bluet{2}\redt{3}\bluet{4}\redt{5} & 
\bluet{1}\bluet{2}\bluet{3}\bluet{4}\bluet{5} & 
\bluet{1}\bluet{2}\bluet{3}\bluet{4}\bluet{5} 
\\

Language (\glpair{L}{})  & \bluet{1}\bluet{2}\redt{3}\bluet{4}\redt{5} & 
\bluet{1}\bluet{2}\redt{3}\bluet{4}\redt{5} & 
- & 
- & 
- 
\\

Year (\glpair{Y}{})  & \redt{1}\redt{2}\redt{3}\redt{4}\redt{5} & 
\redt{1}\redt{2}\redt{3}\redt{4}\redt{5} & 
- & 
\redt{1}\redt{2}\redt{3}\redt{4}\redt{5} & 
- 
\\

Prevalence (\glpair{V}{})  & \redt{1}\bluet{2}\redt{3}\bluet{4}\bluet{5} & 
\bluet{1}\redt{2}\redt{3}\bluet{4}\bluet{5} & 
\redt{1}\redt{2}\bluet{3}\redt{4}\redt{5} & 
- & 
-
\\

\bottomrule
\end{tabular}
}

\end{center}
\end{table}

\subsubsection{Gender} The two groups we consider for gender are ``male'' (G1) and ``female'' (G2), and the data is gathered from MusicBrainz (MB)~\cite{swartz2002musicbrainz}. The experiments we perform for this attribute are denoted as \glpair{G}{Q} (queries from male vs.\ female artists), \glpair{G}{R} (references from male vs.\ female artists), \glpair{G}{C} (compositions from male vs.\ female composers), \glpair{G}{SD} (queries and references from the same vs.\ different gender groups), and \glpair{G}{D12} (queries from male and references from female artists vs.\ queries from female and references from male artists).

\subsubsection{Popularity} For this attribute, we categorize the recordings as ``popular'' (G1) and ``not so popular'' (G2). We consider an artist (whether a person or a band) as popular if they ever had a number-one selling single in the top-10 music markets found in the IFPI’s Global Music Report 2021~\cite{ifpi} (excluding South Korea and China). By including European and Far-eastern music markets, we aim to mitigate bias toward artists from the United States. In the case of composers, we consider them popular if a popular artist (as defined above) ever played a version of any of their compositions. The number-one selling singles data is collected from Wikipedia, and the data for all the compositions of every composer are collected from MB. 
Although there are many ways to define ``popularity,'' our decision was based on the fact that number-one selling single data is the only data that can be found consistently (as opposed to, for example, top-10 singles) for all the aforementioned music markets on Wikipedia, which in itself suggests a distinction between the number-one selling artists and the rest.
The experiments we perform for this attribute are denoted as \glpair{P}{Q}, \glpair{P}{R}, \glpair{P}{C}, \glpair{P}{SD}, and \glpair{P}{D12}.

\subsubsection{Country} For studying the effect of the country, we divide recordings by the ones whose artists or composers are from the United States or the United Kingdom (G1), and the rest (G2), as the songs in our dataset consist of mostly genres originated in music traditions of those countries. In MB, there exist two annotations for the country: ``area,'' the place where the artist currently resides, and the ``begin-area,'' the place where the artist is originally from. In cases where an artist has both annotations, we use the begin-area for our analysis. The experiments we perform for this attribute are denoted as \glpair{C}{Q}, \glpair{C}{R}, \glpair{C}{C}, \glpair{C}{SD}, and \glpair{C}{D12}.

\subsubsection{Language} The two categories we consider for the language are ``English'' (G1) and other (G2). The language data is gathered from SecondHandSongs.com (SHS). The experiments for this attribute are denoted as \glpair{L}{Q} and \glpair{L}{R}. We exclude the other experiments since we cannot obtain the data for the language of compositions (\glpair{L}{C}) and we do not have enough samples for analyzing \glpair{L}{SD} and \glpair{L}{D12}.

\subsubsection{Year} For this attribute, we divide the recordings into two groups by whether they are from the year 2000 or after (G1) or from before (G2). Although this threshold can be set in many different ways, we use the year 2000 as the peer-to-peer networks that changed the music ecosystem drastically became popular by then. The release year data is gathered from SHS. For this attribute, the experiments we perform are denoted as \glpair{Y}{Q} and \glpair{Y}{R}. We also analyze the system performances in cases where the year gap between the query and the reference recordings is less than 10 years (G1) vs.\ greater than or equal to 10 years (G2) and denote the experiment as \glpair{Y}{SD}.

\begin{table}[tb!]
\caption{Sample sizes per group per experiment.}\label{tab:sample-size}
\begin{center}
\begin{tabular}{l r r | l r r}
\toprule
Exp. & \multicolumn{1}{c}{G1} & \multicolumn{1}{c|}{G2} & Exp. & \multicolumn{1}{c}{G1} & \multicolumn{1}{c}{G2} \\
\midrule

\glpair{G}{Q} & 8,067 & 3,296 & \glpair{C}{C} & 13,200 & 3,535  \\
\glpair{G}{R} & 11,976 & 2,954 & \glpair{C}{SD} & 424 & 109  \\
\glpair{G}{C} & 16,840 & 1,093 & \glpair{C}{D12} & 26 & 83 \\
\glpair{G}{SD} & 5,943 & 2,040 & \glpair{L}{Q} & 18,765 & 2,069  \\
\glpair{G}{D12} & 603 & 1,437 & \glpair{L}{R} & 19,487 & 2,172  \\
\glpair{P}{Q} & 2,964 & 17,200 & \glpair{Y}{Q} & 9,310 & 12,019  \\
\glpair{P}{R} & 6,432 & 13,505 & \glpair{Y}{R} & 2,579 & 18,517  \\
\glpair{P}{C} & 3,171 & 1,255 & \glpair{Y}{SD} & 7,737 & 11,129  \\
\glpair{P}{SD} & 11,638 & 6,360 & \glpair{V}{Q} & 11,025 & 11,342  \\
\glpair{P}{D12} & 1,608 & 4,752 & \glpair{V}{R} & 10,211 & 9,882  \\
\glpair{C}{Q} & 1,879 & 638 & \glpair{V}{C} & 9,024 & 13,388  \\
\glpair{C}{R} & 2,967 & 594 & & &  \\

\bottomrule
\end{tabular}

\end{center}
\end{table}

\subsubsection{Prevalence} This last attribute simply characterizes either how prevalent it is for a composition to have a version, quantified by the number of versions a composition has, or how prevalent it is for an artist to perform versions of other artists’ songs, quantified by the number of versions an artist performed. 
We include this attribute in our analyses to see whether songs that have been played by many artists (or artists that have played songs from many others) are in an advantageous position in terms of DRM or not.
The data we use to form the categories are collected from SHS. For the artist prevalence experiments (\glpair{V}{Q} and \glpair{V}{R}), the two groups we consider are the artists who performed less than or equal to 25~versions (G1) and more than 25~versions (G2). For the composition prevalence experiment (\glpair{V}{C}), the two groups we consider are the compositions that have less than or equal to 5 versions (G1) and more than 5 versions (G2). Thresholds 25 and 5 are chosen as they are the median values of their respective distributions.

\input{X_results_table}

\subsection{Dataset and Evaluation}
\subsubsection{Dataset}\label{sssec:dataset} We perform our analyses using a subset of the SHS4- dataset~\cite{doras2019cover}, for which we populated the metadata and label annotations. We publicly release this subset under the name \datasetname, with Creative Commons BY-NC-SA 4.0 license~\cite{cc4.0}. 
The reference set of \datasetname\ includes only one recording per composition, and we specifically chose them as the original recordings for all the compositions (as stated in SHS). Therefore, every query has exactly one correct item in the reference set. The reason for this decision is simply to imitate industrial cases where it is common to have only the original recording in the reference set. However, there can be multiple queries that have the same reference recording as the correct item. The total number of queries and references in \datasetname\ is 22,428 and 15,293, respectively. 
All the deep learning-based models are trained using the SHS5+ dataset~\cite{doras2019cover}. For the training details of the considered systems, see Yesiler et al.~\cite{yesiler2019accurate} and Doras et al.~\cite{doras2020combining}.

\subsubsection{Evaluation} We assess system performances using the reciprocal rank metric, which we denote by $\psi$. We motivate this choice by the fact that $\psi$ penalizes the differences in lower ranks more than the differences in higher ranks, which we consider as a good proxy for real use cases (e.g.,~the difference between the correct answer being at rank 1 or 11 should have more impact than the difference of it being at rank 31 or 41). Note that, since there is only one correct item in the reference collection for each query, $\psi$ corresponds to the same score as average precision. 

For each group in our study, we collect $\psi$ scores for all the items belonging to that group and use them to form performance distributions. To compare the obtained distributions of two groups, we use the two-sample Kolmogorov-Smirnov (KS) test~\cite{kstest}, which has several characteristics that are desirable for our setup (e.g.,~it is a non-parametric exact test, and it is not sensitive to imbalanced sample sizes between groups). The null hypothesis of the KS test is that the samples are drawn from the same distribution (i.e.,~that the underlying distribution is the same). To reject the null hypothesis, we consider the threshold of 0.05 for the $p$-value; however, since we test multiple hypotheses, we apply the Holm-Bonferroni correction~\cite{holmcorrect}. For computing the KS test and plotting rank distributions (see Figure~\ref{fig:move}), we collapse all the ranks above 10, to remove potential differences in the tails (i.e.,~high ranks that, with high probability, are not going to be considered by practitioners). For each analysis, we report sample sizes (see Table~\ref{tab:sample-size}) and the mean $\psi$ scores, $\bar{\psi}$, for each group.

%% file: X_results_table.tex
\begin{table*}[htp!]
\caption{Detailed results. 
For each system and experiment, we report the raw $p$-value of the KS test and the mean reciprocal rank of groups 1 and 2 (\mrrsym-G1 and \mrrsym-G2). \redt{Red} color with \sigone\ denotes statistical significance after the Holm-Bonferroni correction. Overall system performance with the considered dataset is indicated in parenthesis in the top row.
}\label{tab:all-results}
\begin{center}
\begin{tabular}{l | ccc | ccc | ccc | ccc | ccc}
\toprule 
 & \multicolumn{3}{c|}{Qmax (\mrrsym=0.42)} &  \multicolumn{3}{c|}{MOVE (\mrrsym=0.63)} & \multicolumn{3}{c|}{MICE-F (\mrrsym=0.50)} & \multicolumn{3}{c|}{MICE-C (\mrrsym=0.59)} &  \multicolumn{3}{c}{LF-c (\mrrsym=0.75)} \\
Exp. & \multicolumn{1}{c}{$p$} & \multicolumn{1}{c}{\mrrsym-G1} & \multicolumn{1}{c|}{\mrrsym-G2} & \multicolumn{1}{c}{$p$} & \multicolumn{1}{c}{\mrrsym-G1} & \multicolumn{1}{c|}{\mrrsym-G2} & \multicolumn{1}{c}{$p$} & \multicolumn{1}{c}{\mrrsym-G1} & \multicolumn{1}{c|}{\mrrsym-G2} & \multicolumn{1}{c}{$p$} & \multicolumn{1}{c}{\mrrsym-G1} & \multicolumn{1}{c|}{\mrrsym-G2} & \multicolumn{1}{c}{$p$} & \multicolumn{1}{c}{\mrrsym-G1} & \multicolumn{1}{c}{\mrrsym-G2} \\
\midrule

\glpair{G}{Q} & 
<0.01 & 0.41 & 0.45 & \redt{<0.01}\sigone & 0.63 & 0.70 & 
\redt{<0.01}\sigone & 0.46 & 0.61 &
\redt{<0.01}\sigone & 0.58 & 0.65 &
\redt{<0.01}\sigone & 0.73 & 0.81  \\
\glpair{G}{R}& 
0.41 & 0.42 & 0.44  & \redt{<0.01}\sigone & 0.63 & 0.69 & 
\redt{<0.01}\sigone & 0.48 & 0.64 &
\redt{<0.01}\sigone & 0.59 & 0.65 &
\redt{<0.01}\sigone & 0.74 & 0.83 \\
\glpair{G}{C} & 
0.19 & 0.42 & 0.39 & <0.01 & 0.63 & 0.67 & 
\redt{<0.01}\sigone & 0.49 & 0.60 &
<0.01 & 0.58 & 0.63 &
\redt{<0.01}\sigone & 0.74 & 0.80 \\
\glpair{G}{SD} & 
0.14 & 0.44 & 0.42 & \redt{<0.01}\sigone & 0.65 & 0.70 & 
\redt{<0.01}\sigone & 0.51 & 0.56 &
\redt{<0.01}\sigone & 0.60 & 0.65 &
\redt{<0.01}\sigone & 0.74 & 0.81 \\
\glpair{G}{D12} & 
0.05 & 0.38 & 0.44 & 1.00 & 0.70 & 0.71 & 
0.09 & 0.52 & 0.57 &
1.00 & 0.65 & 0.65 &
1.00 & 0.80 & 0.81 \\
\hline
\glpair{P}{Q} & 
0.06 & 0.44 & 0.42 & \redt{<0.01}\sigone & 0.71 & 0.62 & 
\redt{<0.01}\sigone & 0.58 & 0.49 &
\redt{<0.01}\sigone & 0.66 & 0.57 &
\redt{<0.01}\sigone & 0.82 & 0.74 \\
\glpair{P}{R} & 
\redt{<0.01}\sigone & 0.47 & 0.40 & \redt{<0.01}\sigone & 0.70 & 0.60 & 
\redt{<0.01}\sigone & 0.55 & 0.48 &
\redt{<0.01}\sigone & 0.66 & 0.56 &
\redt{<0.01}\sigone & 0.80 & 0.72 \\
\glpair{P}{C} & 
0.57 & 0.41 & 0.39 & <0.01 & 0.63 & 0.58 & 
0.83 & 0.50 & 0.48 &
0.01 & 0.59 & 0.54 &
0.38 & 0.75 & 0.73 \\
\glpair{P}{SD} & 
\redt{<0.01}\sigone & 0.41 & 0.45 & \redt{<0.01}\sigone & 0.61 & 0.69 & 
\redt{<0.01}\sigone & 0.48 & 0.53 &
\redt{<0.01}\sigone & 0.57 & 0.64 &
\redt{<0.01}\sigone & 0.73 & 0.79 \\
\glpair{P}{D12} & 
0.02 & 0.43 & 0.46  & 1.00 & 0.69 & 0.69 & 
0.03 & 0.55 & 0.53 &
0.99 & 0.63 & 0.64 &
1.00 & 0.79 & 0.79 \\
\hline
\glpair{C}{Q} & 
0.96 & 0.42 & 0.41 & 1.00 & 0.61 & 0.61 & 
0.98 & 0.43 & 0.45 &
0.78 & 0.55 & 0.57 &
0.54 & 0.69 & 0.73 \\
\glpair{C}{R} & 
0.72 & 0.43 & 0.44 & 0.34 & 0.61 & 0.65 & 
0.78 & 0.43 & 0.45 &
0.02 & 0.57 & 0.63 &
0.89 & 0.71 & 0.73 \\
\glpair{C}{C} & 
0.62 & 0.41 & 0.42 & <0.01 & 0.63 & 0.65 & 
\redt{<0.01}\sigone & 0.49 & 0.56 &
<0.01 & 0.58 & 0.61 &
\redt{<0.01}\sigone & 0.74 & 0.79 \\
\glpair{C}{SD} & 
1.00 & 0.45 & 0.47 & 0.30 & 0.65 & 0.58 & 
0.88 & 0.42 & 0.39 &
0.63 & 0.60 & 0.56 &
0.66 & 0.72 & 0.69 \\
\glpair{C}{D12} & 
0.66 & 0.59 & 0.44 & 0.22 & 0.73 & 0.54 & 
1.00 & 0.40 & 0.39 &
0.54 & 0.67 & 0.53 &
1.00 & 0.69 & 0.69 \\
\hline
\glpair{L}{Q} & 
1.00 & 0.42 & 0.41 & 0.07 & 0.64 & 0.66 & 
\redt{<0.01}\sigone & 0.51 & 0.64 &
0.12 & 0.59 & 0.61 &
\redt{<0.01}\sigone & 0.75 & 0.82 \\
\glpair{L}{R} & 
0.98 & 0.42 & 0.41 & 0.06 & 0.64 & 0.66 & 
\redt{<0.01}\sigone & 0.50 & 0.63 &
0.07 & 0.59 & 0.61 &
\redt{<0.01}\sigone & 0.75 & 0.82 \\
\hline
\glpair{Y}{Q} & 
\redt{<0.01}\sigone & 0.46 & 0.39 & \redt{<0.01}\sigone & 0.66 & 0.61 & 
\redt{<0.01}\sigone & 0.52 & 0.48 &
\redt{<0.01}\sigone & 0.62 & 0.56 &
\redt{<0.01}\sigone & 0.77 & 0.73 \\
\glpair{Y}{R} & 
\redt{<0.01}\sigone & 0.48 & 0.41 & \redt{<0.01}\sigone & 0.67 & 0.63 & 
\redt{<0.01}\sigone & 0.64 & 0.48 &
\redt{<0.01}\sigone & 0.65 & 0.58 &
\redt{<0.01}\sigone & 0.82 & 0.73 \\
\glpair{Y}{SD} & 
\redt{<0.01}\sigone & 0.44 & 0.40 & \redt{<0.01}\sigone & 0.67 & 0.60 & 
\redt{<0.01}\sigone & 0.53 & 0.45 &
\redt{<0.01}\sigone & 0.63 & 0.55 &
\redt{<0.01}\sigone & 0.78 & 0.71 \\
\hline
\glpair{PR}{Q} & 
\redt{<0.01}\sigone & 0.43 & 0.40 & 0.16 & 0.62 & 0.64 & 
\redt{<0.01}\sigone & 0.49 & 0.52 &
0.37 & 0.58 & 0.59 &
0.18 & 0.74 & 0.75 \\
\glpair{PR}{R} & 
0.11 & 0.43 & 0.41 & \redt{<0.01}\sigone & 0.62 & 0.65 & 
\redt{<0.01}\sigone & 0.49 & 0.52 &
0.04 & 0.58 & 0.60 &
<0.01 & 0.74 & 0.76 \\
\glpair{PR}{C} & 
\redt{<0.01}\sigone & 0.45 & 0.40 & \redt{<0.01}\sigone & 0.65 & 0.61 & 
1.00 & 0.50 & 0.50 &
\redt{<0.01}\sigone & 0.61 & 0.57 &
\redt{<0.01}\sigone & 0.76 & 0.74 \\

\bottomrule
\end{tabular}

\end{center}
\vspace{-2mm}
\end{table*}

%% file: 3_results.tex
\section{Results}\label{sec:results}

\begin{figure*}[th!]
\includegraphics[width=1\linewidth]{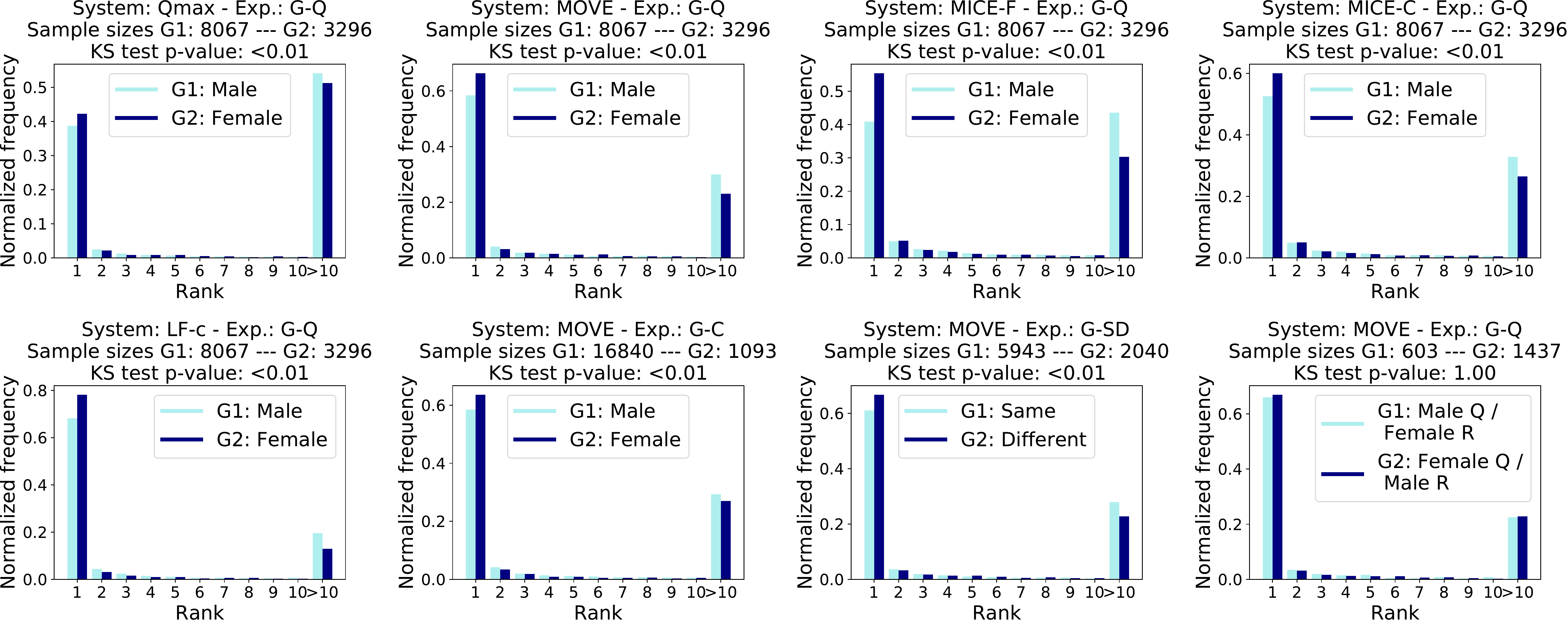}
\caption{Rank distributions for a selected set of experiments. The full set of 115~experiments is available in the appendix.}
\label{fig:move}
\Description[Eight figures of histograms showing rank distributions for G1 and G2 each for a different experiment.]{Eight figures of histograms showing rank distributions for G1 and G2 each for a different experiment. The first four figures compare the rank distributions for queries whose artists are male or female across all five of the considered systems. The system performances appear better for tracks of female artists, indicated by p-values less than 0.01. The last three figures show rank distributions for different sets of experiments that all use the VI system MOVE. Tracks of female composers are identified better than tracks of male composers. Query and reference tracks of different genders are identified better than tracks of the same gender. Lastly, there is no difference in performance for male queries and female references or female queries and male references. }
\end{figure*}

Table~\ref{tab:all-results} presents the detailed list of results for all the experiments and the considered VI systems, and Figure~\ref{fig:move} presents rank distribution illustrations for a selected set of experiments. We include the rank distribution illustrations for all the experiments along with the detailed scores in our repository. We here highlight the main findings with respect to each considered attribute. Possible reasons for the observed disparities are discussed in Section~\ref{ssec:causes}.

\subsubsection{Gender} The results for \glpair{G}{Q}, \glpair{G}{R}, and \glpair{G}{C} suggest that the identification performances of the learning-based models form two distinct distributions for most systems (see Figure~\ref{fig:move}), and that performances are significantly higher for the recordings performed or composed by female artists (G2). On the contrary, for the only rule-based system we consider, the obtained $p$-values for \glpair{G}{Q}, \glpair{G}{R}, and \glpair{G}{C} do not suggest a difference between the performance for male (G1) and female (G2) artists/composers. The results for \glpair{G}{SD} and \glpair{G}{D12} show that the learning-based systems perform better when the gender of the query and the reference artists are different (G2), and that the direction of the change (female reference to male query, or male reference to female query) does not create a significant difference.

\subsubsection{Popularity} Results for the popularity attribute suggest that all the systems perform better for queries and references from popular artists (G1); only for \glpair{P}{Q} using Qmax does not result in a significant difference. However, such a difference between groups is not observed for the composers (\glpair{P}{C}). Moreover, when the artists of the query and the reference tracks belong to different popularity groups (G2), identification performance is higher (\glpair{P}{SD}). As with the case with gender, the direction of the change does not seem to create much of a difference (\glpair{P}{D12}).

\subsubsection{Country} Based on the results for the country attribute, we do not see any differences between performances with respect to the country of the performing artists (\glpair{C}{Q} and \glpair{C}{R}). However, MICE-F and LF-c systems show such a difference with respect to the country of the composer, favoring composers from outside the US/UK (G2) group (\glpair{C}{C}). We also observe no differences for the cases where the query and the reference recordings belong to the same (G1) vs.\ different (G2) groups (\glpair{C}{SD}) and when the direction of the change is different (\glpair{C}{D12}), but note that the sample sizes for the last two experiments are smaller than the previous ones.

\subsubsection{Language} The results for the language attribute suggest a performance difference between groups for the systems that use melody-based features (i.e.,~MICE-F and LF-c) but not for the ones that use only chroma-based features (\glpair{L}{Q} and \glpair{L}{R}).

\subsubsection{Year} All the considered experiments for the year category suggest the same result across all the systems: identification performance for queries and references after 2000 (G1) is higher compared to queries and references before that year (\glpair{Y}{Q} and \glpair{Y}{R}). Also, when the year gap between the query and the reference tracks is less than a decade (G1), the systems perform better.

\subsubsection{Prevalence} We find no pattern behind the differences in the results for query recordings from artists that perform versions of other artists' compositions less (G1) or more (G2) prevalently (\glpair{V}{Q}): while the performances of the considered groups seem similar for MOVE, MICE-C, and LF-c systems, MICE-F and Qmax favor different groups. For the experiments that compare the same quality for reference recordings (\glpair{V}{R}), however, we see that the learning-based systems favor the cases where the reference recordings are from artists that perform versions more prevalently (G2) than the others (G1), while Qmax favors the opposite (G1 over G2). Lastly, in terms of comparing compositions with fewer (G1) or more (G2) versions (\glpair{V}{C}), all the systems except MICE-F suggest a significant difference between the considered groups, favoring compositions with fewer versions (G1) over the others (G2).

%% file: 4_discussion.tex
\section{Discussion}\label{sec:discussion}

Before entering into any discussion of the results, we shall first remind ourselves of the limitations of our data collection process. Firstly, the data sources we use (MB, SHS, and Wikipedia) contain human-annotated data, but we still would like to point out the possibility of human error. Secondly, since we excluded the cases where we could not find a label for a certain attribute, this may have created a potential bias on the data we use. Lastly, although we have implemented many heuristics to obtain the correct data from our sources, we have only performed random checks whether we succeeded on that, rather than checking all the obtained annotations one by one. Therefore, there may have been a small amount of metadata matching problems. However, we assume that, even if such cases exist, they should not drastically affect our results. 
In addition to the data collection process, the limitations of creating binary groups may also have an impact on the results. For example, creating such binary groups based on other criteria, or using non-binary partitions might have changed the outcomes of the presented analyses.
Keeping these in mind, we now present our hypotheses on the causes of the observed disparities between groups and discuss the potential fairness implications of the results.

\subsection{Causes of disparities}\label{ssec:causes}

Overall, we observe that the learning-based systems work better for underrepresented groups (see, for instance,~\glpair{G}{Q}, \glpair{G}{R}, \glpair{P}{Q}, and \glpair{L}{Q}). Our first hypothesis for this is that the groups with fewer samples may show less variety between versions in terms of the musical characteristics; thus, the identification performance may end up higher. A second reason for this may be the sample sizes of the groups in the training set, but we see that, for example, the gender ratio (G1 vs.\ G2) in the training set is 2.5, which is not very different than the one of the analysis set, which is 3.2.

When analyzing the attributes separately, we observe that the performances of the learning-based systems are higher for female artists and composers, but also the differences between \mrrsym\ for G1 and G2 are higher for the systems that use melody features. We think that this may be related to the differences between male and female voices and the ability of the melody extraction algorithm to correctly estimate them. Also, looking at the experiments where the query and the reference recordings belong to the same vs.\ different groups for gender and popularity attributes (\glpair{G}{SD} and \glpair{P}{SD}), we see that the learning-based systems perform better for the cases where they belong to different groups (G2). We think that when the groups of the query and the reference change, the resulting version may appear more faithful to the original, and this may positively affect system performance.

For the popularity experiments, we see that all the systems tend to perform better for popular artists and composers (G1), though not all the experiments show a significant difference (\glpair{P}{Q} and \glpair{P}{C}). We hypothesize that this may be due to the input feature extraction algorithms, and the fact that they may have been optimized using tracks from such popular artists. Especially among the learning-based systems, MOVE and MICE-C show larger disparities between groups, and the chord estimation model used for cremaPCP was trained using a dataset including mainly popular tracks.

The results for the language experiments show disparities only for the systems using melody features, with recordings in languages other than English (G2) having better results than recordings in English (G1). A couple of reasons could be that, as mentioned earlier, having fewer samples may cause less variety between versions, and that the phonetic qualities of languages may affect the success of melody estimation algorithms (for example, Italian is argued to be the most suitable language for opera~\cite{italian}). 

In terms of the year attribute, we see that all the systems perform better for queries and references released in the year 2000 or after (G1). Possible reasons for this include the trends in the music creation process, which may affect the degree of variance between versions of a composition; the fact that G1 spans over a narrower window of years, which may naturally limit the possibility of having more different versions; and having fewer samples for G1 (as discussed above). For the year gap experiments (\glpair{Y}{SD}), we argue that when the year gap between versions is longer (G2), it becomes more probable to see versions that are less faithful to the original track, which may cause lower system performances.

\subsection{Considerations for fairness implications}
Fairness evaluations are generally applied to systems that already interact with humans, where their direct impact on the considered groups is studied.
In our case, however, we do not have access to a deployed VI system nor any real metrics on how the considered groups are affected (e.g.,~real amount of circulated royalties). Therefore, we now discuss the ``potential'' implications of the results presented in Section~\ref{sec:results}, rather than evaluating a real-life scenario.
Although the following considerations are speculative in nature, they are based on our quantitative analyses, and we believe that not having access to a real-life system should not preclude research on such problems.

For interpreting the performance differences as fairness outcomes, we take the application of VI in DRM. 
In our multi-stakeholder scenario, we assume a direct connection between identification performance and the financial impact on various parties: if a query is not correctly identified, the performing artist of the query does not pay and the composer does not receive due royalties.
To that extent, the reciprocal rank metric may have opposite meanings for the involved parties. The artist of the query may want a lower identification performance while the composer may desire the opposite.

To better illustrate this point, we can take a look at the gender experiments. We observe that the learning-based systems work better for female artists/composers compared to males. While this implies that female composers are likely to be more rewarded by these systems, in contrast, female artists that perform a version of an existing composition (i.e.,~artists of the queries) are likely to pay more royalties. Therefore, we here have a case where interpreting the fairness outcomes should be considered independently for all the involved parties, as a result of having a multi-sided structure.

Lastly, we observe that both the learning- and the rule-based systems show performance disparities for certain groups. Specifically, the learning-based systems show disparities for 54.4\% of the cases while this ratio is only 30.4\% for the rule-based system. While this clearly presents a disadvantage for the learning-based systems, their advantages regarding accuracy and scalability make them irreplaceable for the industry. Therefore, to make sure such systems do not create any unfair conditions that favor certain groups, evaluating VI systems that may have a real-life impact should incorporate fairness and algorithmic bias metrics, along with metrics for measuring accuracy and scalability. 

%% file: 5_conclusion.tex
\section{Conclusion}
In this work, we have investigated the disparities in the performance of VI systems that may have an effect on fairness regarding the involved stakeholders. We have used 6~attributes to create potentially impacted groups based on the performing artists of the query, the performing artists of the reference, and the composer of the composition, which we identify as the related parties in a VI workflow. For our analyses, we have collected annotations for the attributes we consider and created \datasetname\ dataset, which we publicly share. As the result of 115~experiments in total, we have seen that VI systems may indeed perform differently on certain groups. Their behavior may vary depending on whether they are learning- or rule-based, or whether they use melody- or chroma-based input features, but potentially other design choices could have an impact. After presenting the obtained results, we have proposed our hypotheses on the possible causes of such performance disparities. We leave confirming or rejecting them as future work. Lastly, we have discussed the potential implications of the obtained results on multiple stakeholders involved in VI from a fairness perspective. We encourage future VI research to incorporate fairness- and algorithmic bias--related evaluation metrics along with the existing accuracy- and scalability-related ones, to get ahold of any wrongful practices.

%% file: 6_appendix.tex
\clearpage\onecolumn
\appendix
\renewcommand{\thetable}{S\arabic{table}}  
\renewcommand{\thefigure}{S\arabic{figure}}
\setcounter{figure}{0}
\setcounter{table}{0}
\newcommand{\vvspace}{\vspace{-9mm}}

\section{Rank distributions and detailed results}
\subsection{Gender experiments}

\begin{figure*}[h!]
\includegraphics[width=1\linewidth]{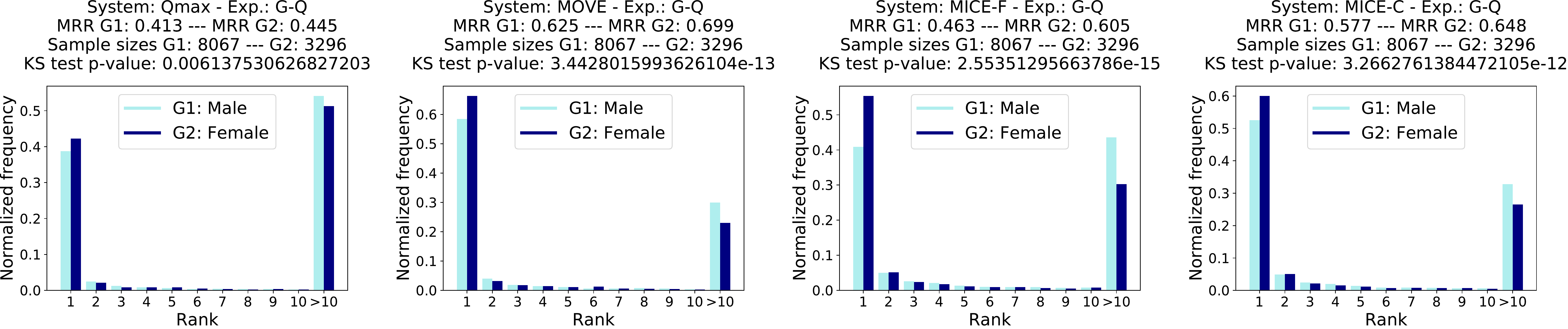}
\vvspace
\end{figure*}

\begin{figure*}[h!]
\includegraphics[width=1\linewidth]{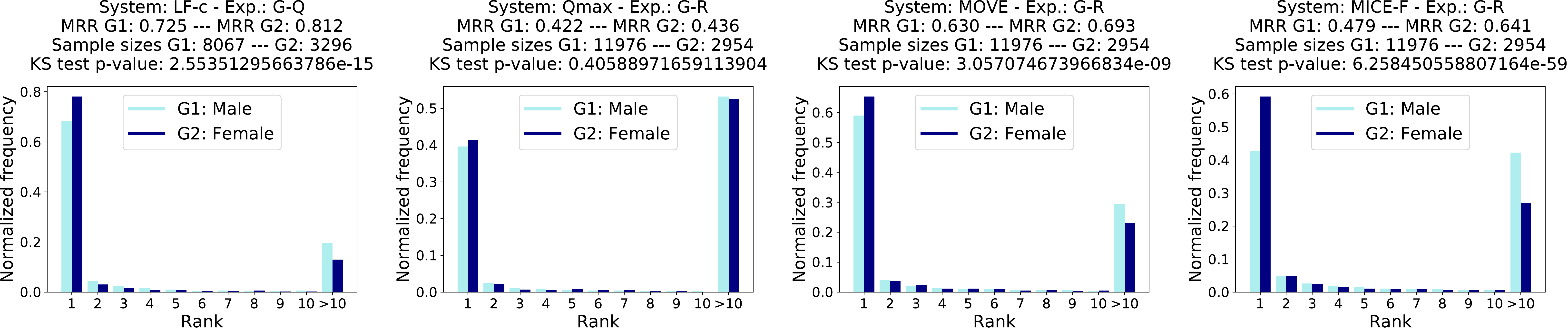}
\vvspace
\end{figure*}

\begin{figure*}[h!]
\includegraphics[width=1\linewidth]{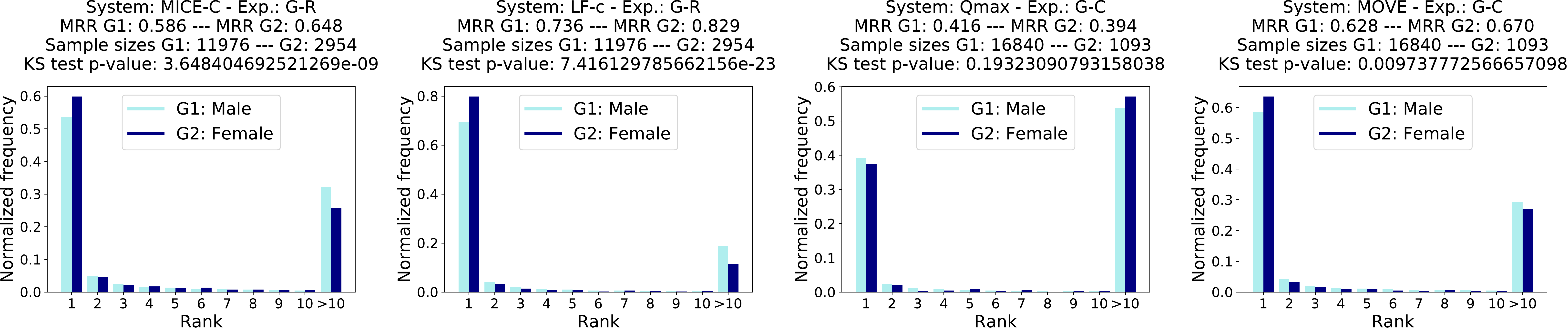}
\vvspace
\end{figure*}

\begin{figure*}[h!]
\includegraphics[width=1\linewidth]{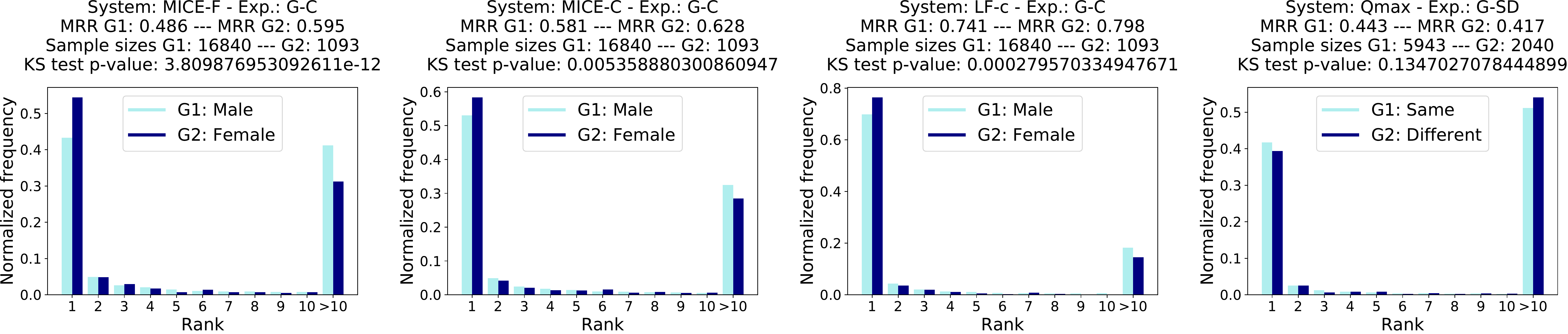}
\vvspace
\end{figure*}

\begin{figure*}[h!]
\includegraphics[width=1\linewidth]{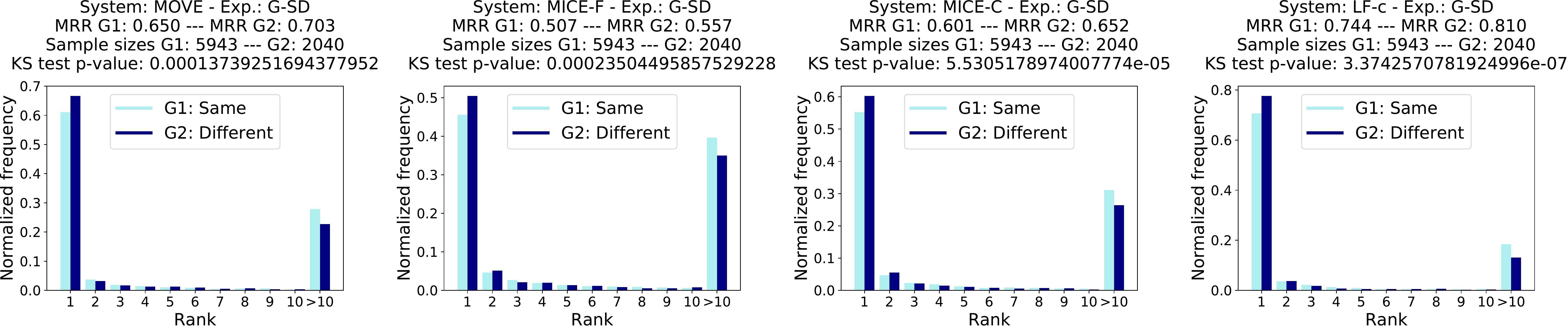}
\end{figure*}

\begin{figure*}[h!]
\includegraphics[width=1\linewidth]{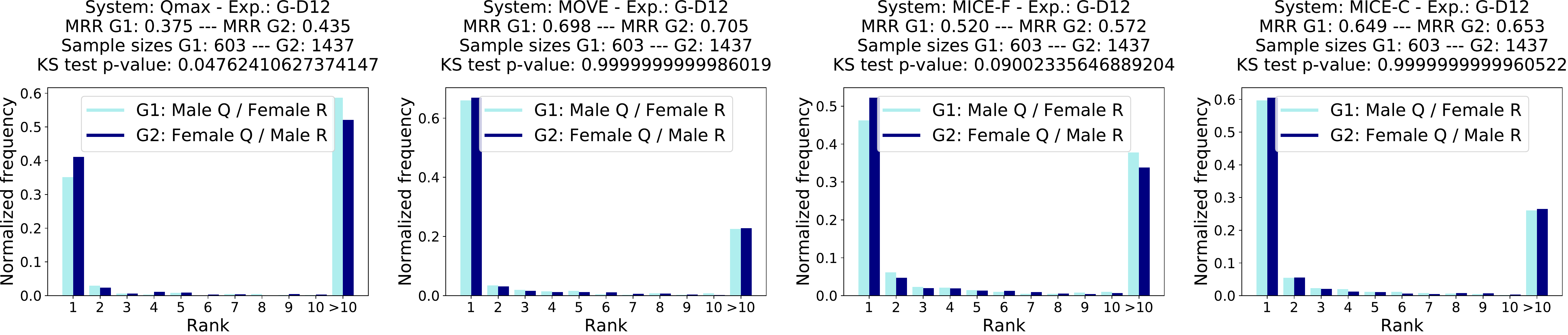}
\end{figure*}

\begin{figure*}[h!]
\includegraphics[width=1\linewidth]{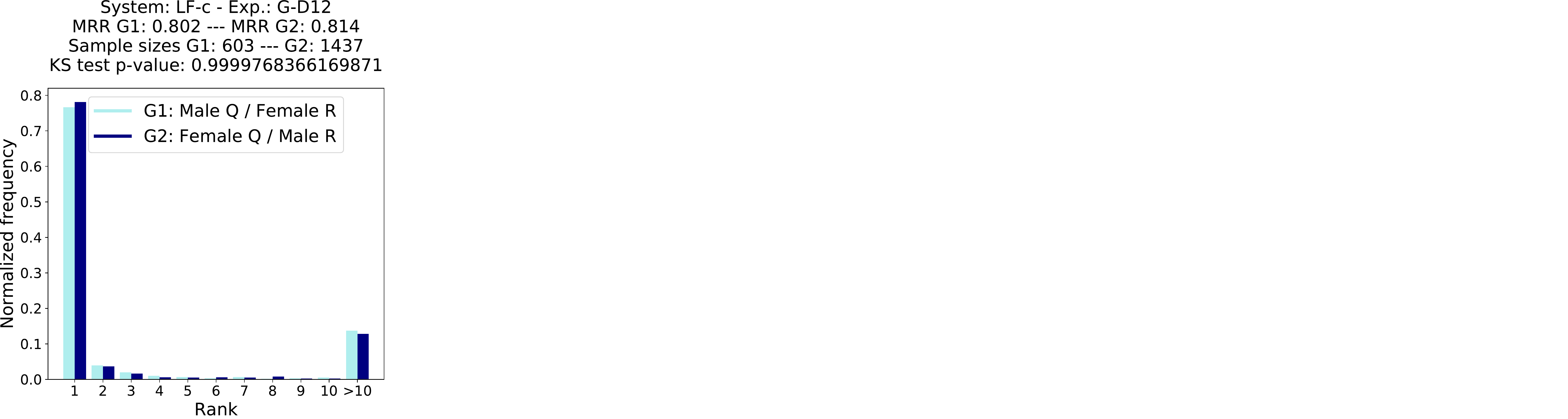}
\end{figure*}

\newpage
\subsection{Population experiments}

\begin{figure*}[h!]
\includegraphics[width=1\linewidth]{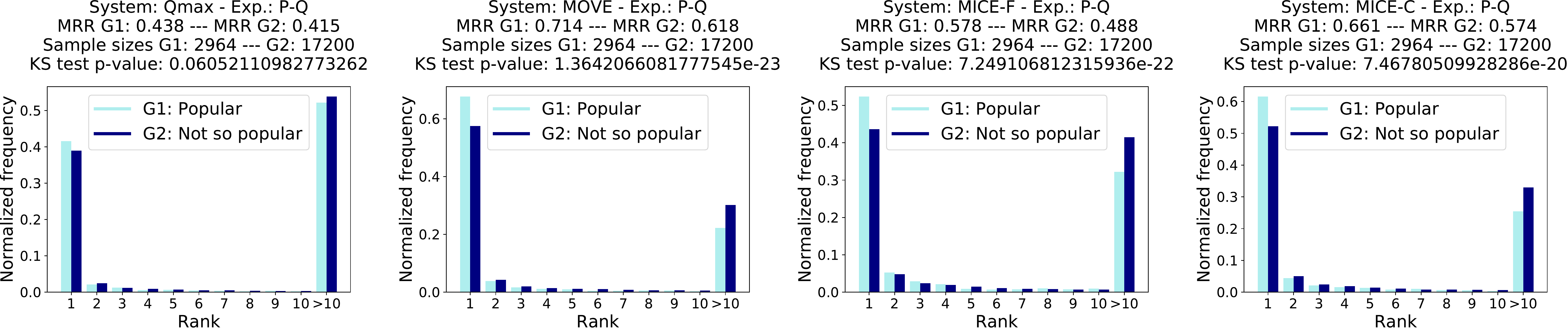}
\vvspace
\end{figure*}

\begin{figure*}[h!]
\includegraphics[width=1\linewidth]{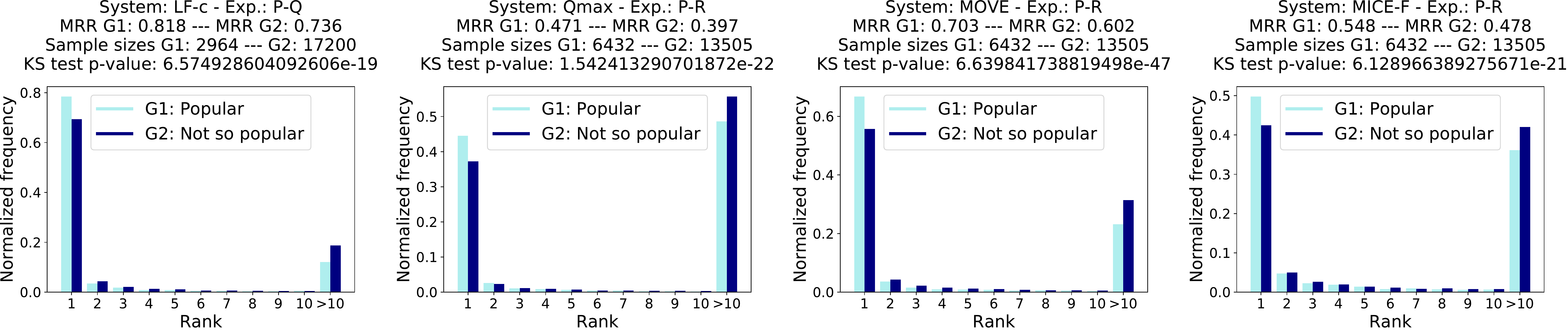}
\vvspace
\end{figure*}

\begin{figure*}[h!]
\includegraphics[width=1\linewidth]{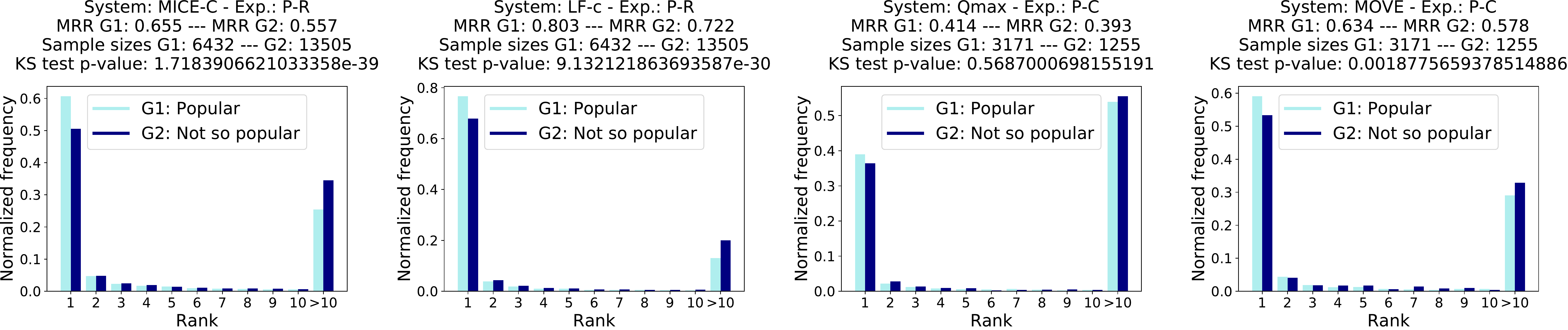}
\vvspace
\end{figure*}

\begin{figure*}[h!]
\includegraphics[width=1\linewidth]{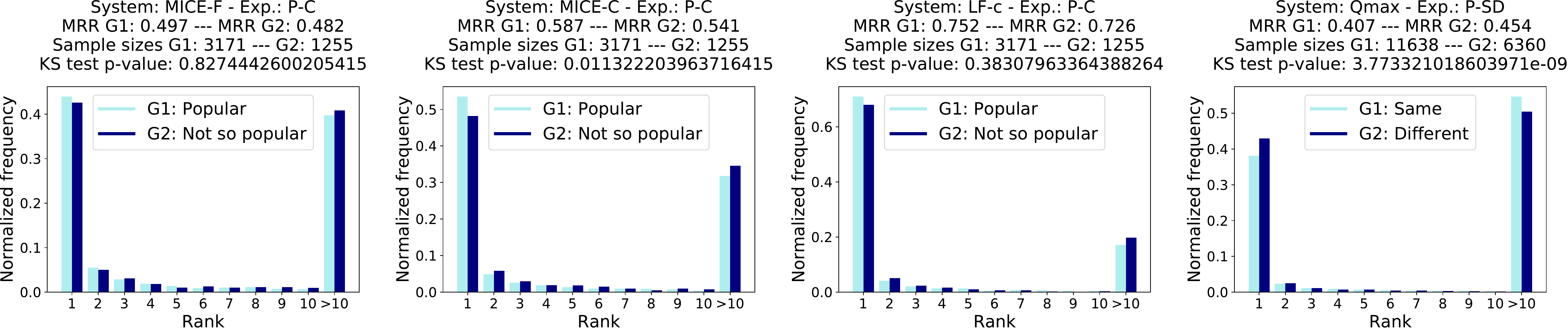}
\end{figure*}

\begin{figure*}[h!]
\includegraphics[width=1\linewidth]{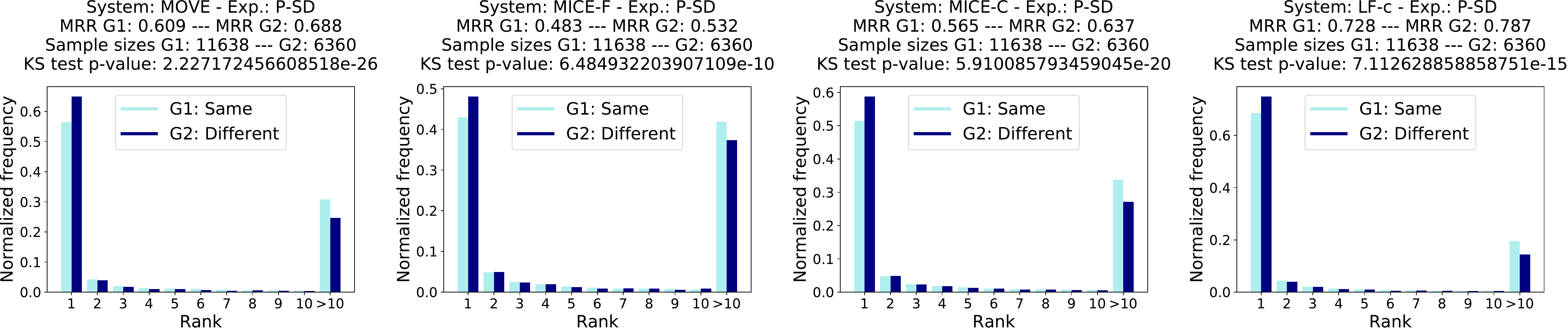}
\end{figure*}

\begin{figure*}[h!]
\includegraphics[width=1\linewidth]{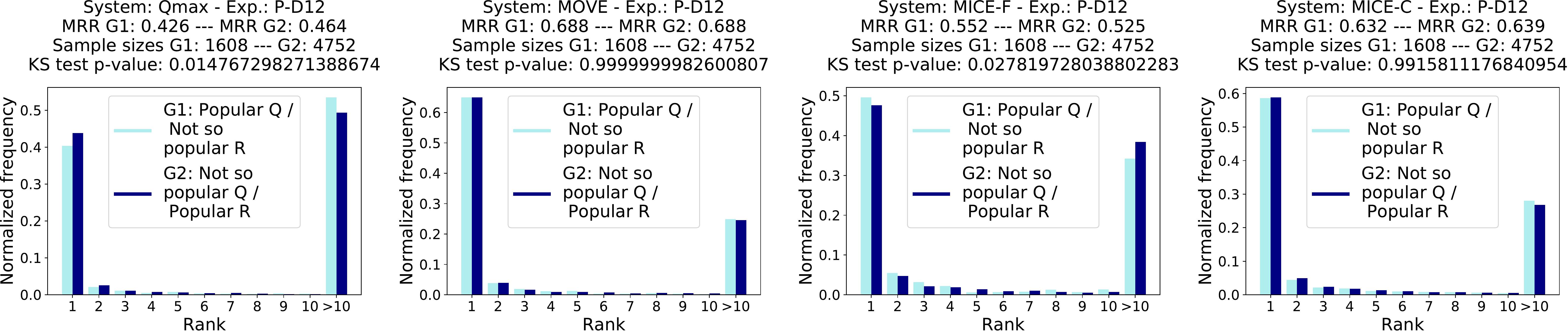}
\end{figure*}

\begin{figure*}[h!]
\includegraphics[width=1\linewidth]{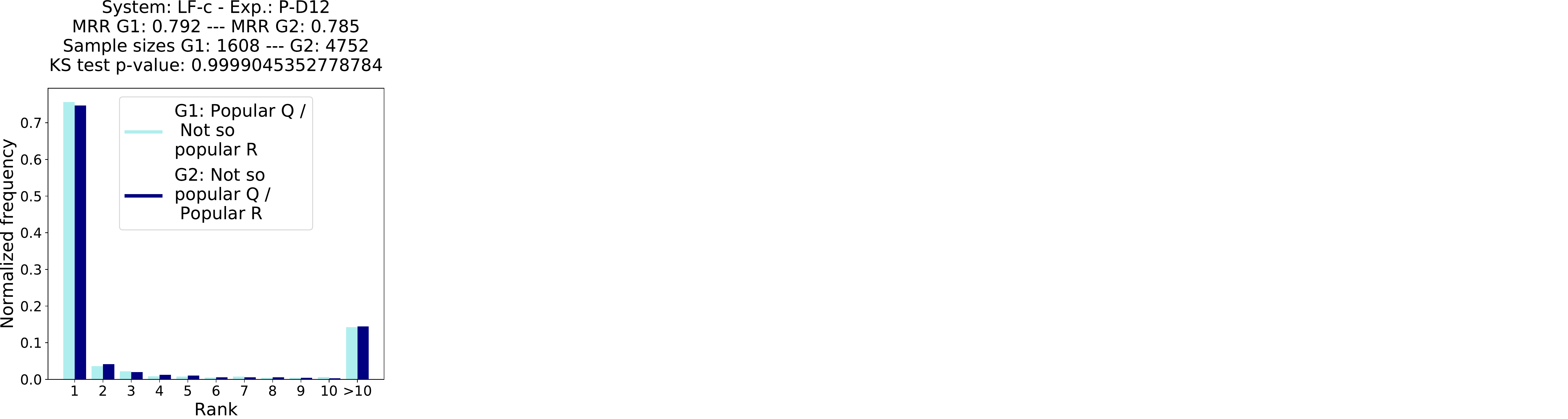}
\end{figure*}

\clearpage
\subsection{Country experiments}

\begin{figure*}[h!]
\includegraphics[width=1\linewidth]{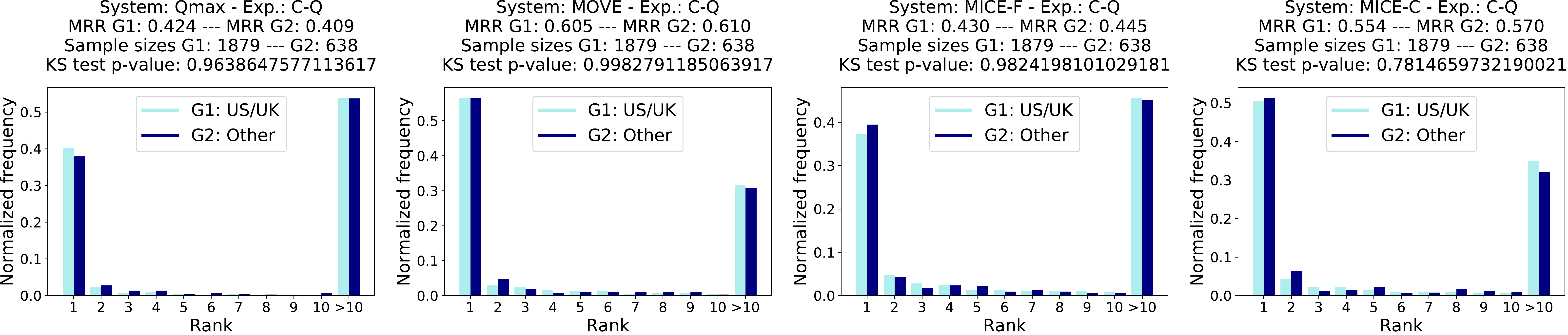}
\vvspace
\end{figure*}

\begin{figure*}[h!]
\includegraphics[width=1\linewidth]{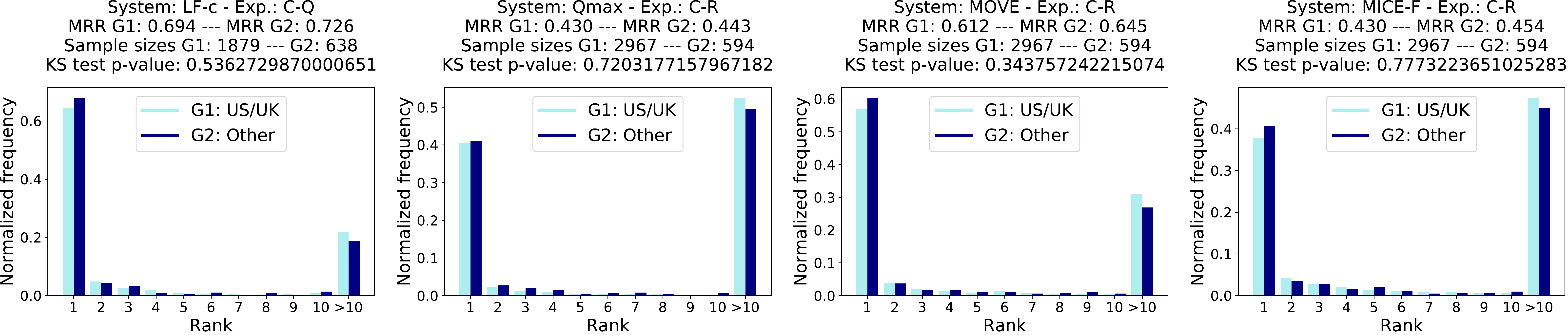}
\vvspace
\end{figure*}

\begin{figure*}[h!]
\includegraphics[width=1\linewidth]{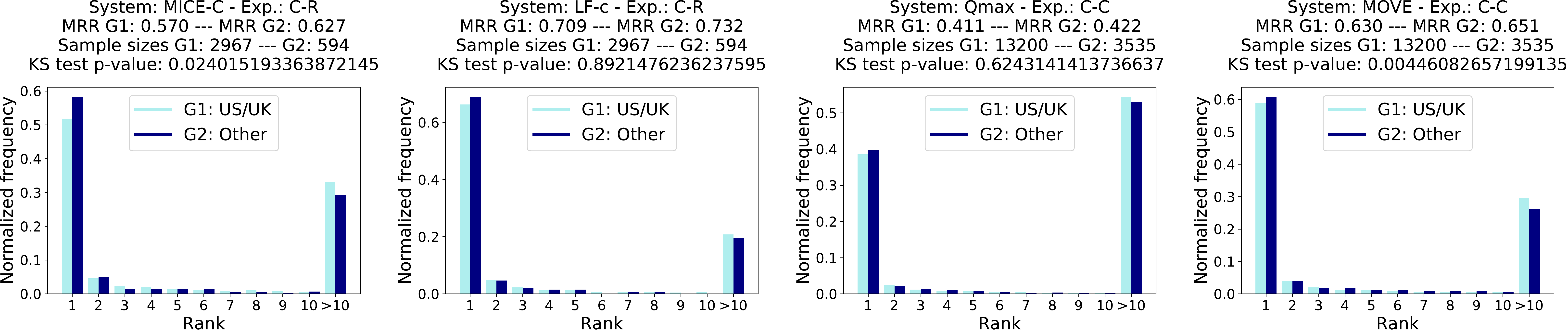}
\vspace{-7mm}
\end{figure*}

\begin{figure*}[h!]
\includegraphics[width=1\linewidth]{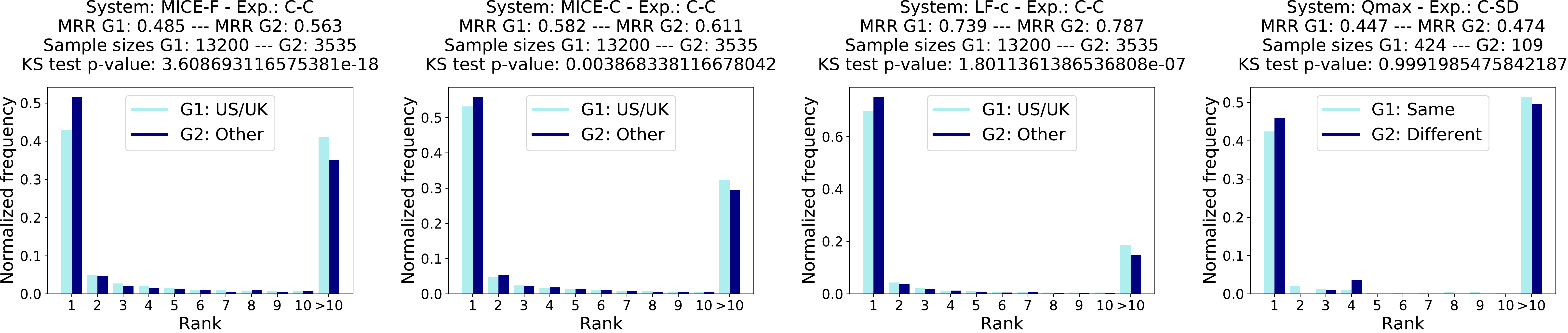}
\vspace{-7mm}
\end{figure*}

\begin{figure*}[h!]
\includegraphics[width=1\linewidth]{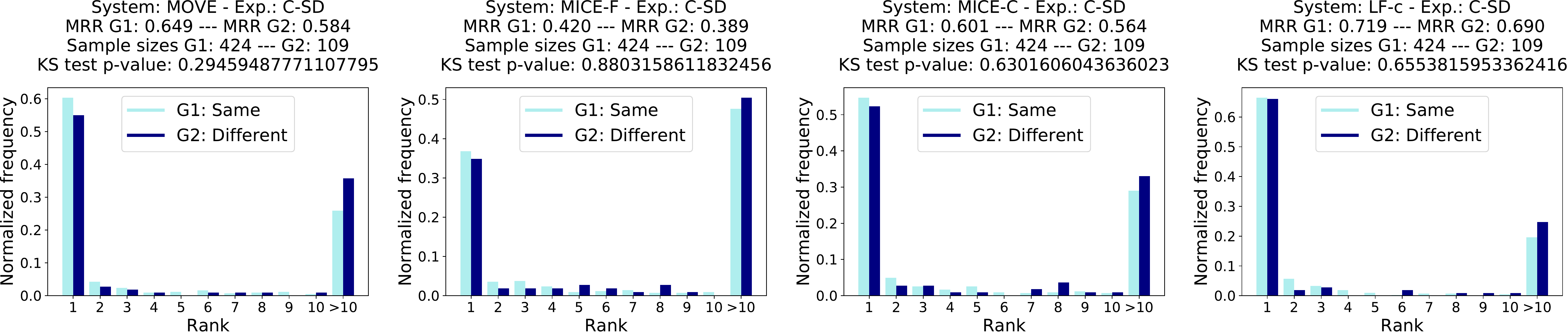}
\vspace{-7mm}
\end{figure*}

\begin{figure*}[h!]
\includegraphics[width=1\linewidth]{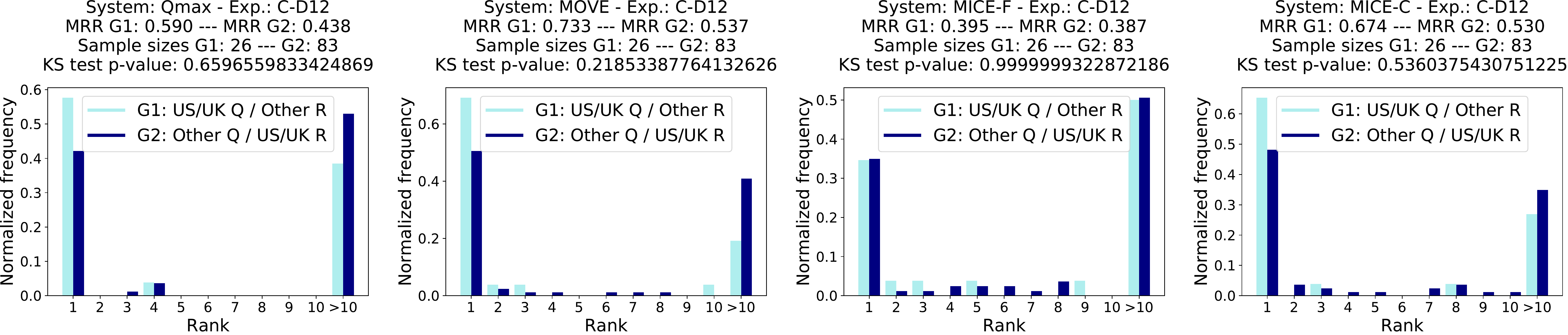}
\vspace{-7mm}
\end{figure*}

\begin{figure*}[h!]
\includegraphics[width=1\linewidth]{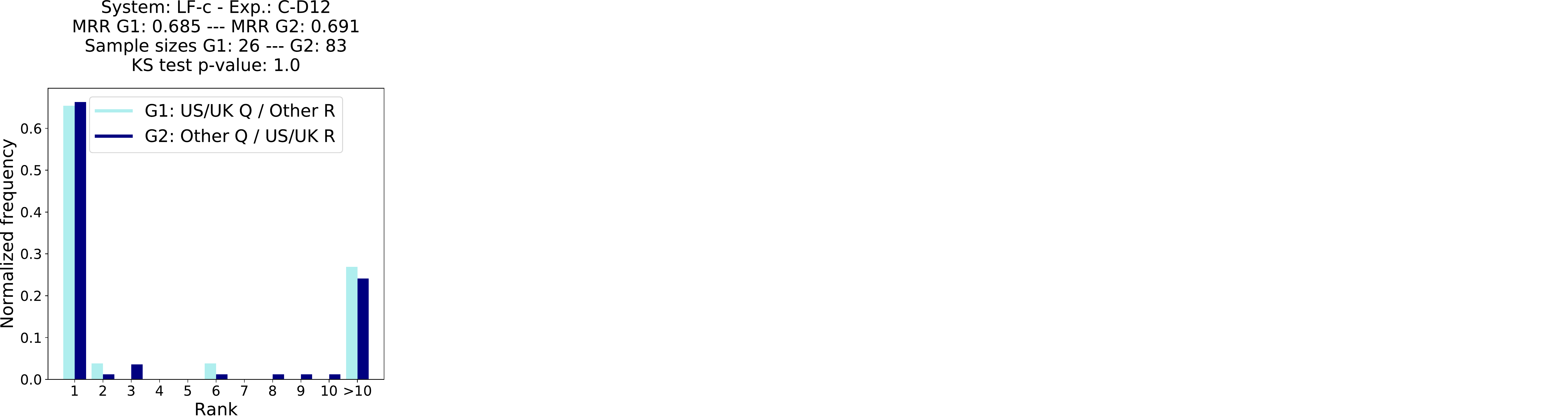}
\vspace{-7mm}
\end{figure*}

\newpage
\subsection{Language experiments}

\begin{figure*}[h!]
\includegraphics[width=1\linewidth]{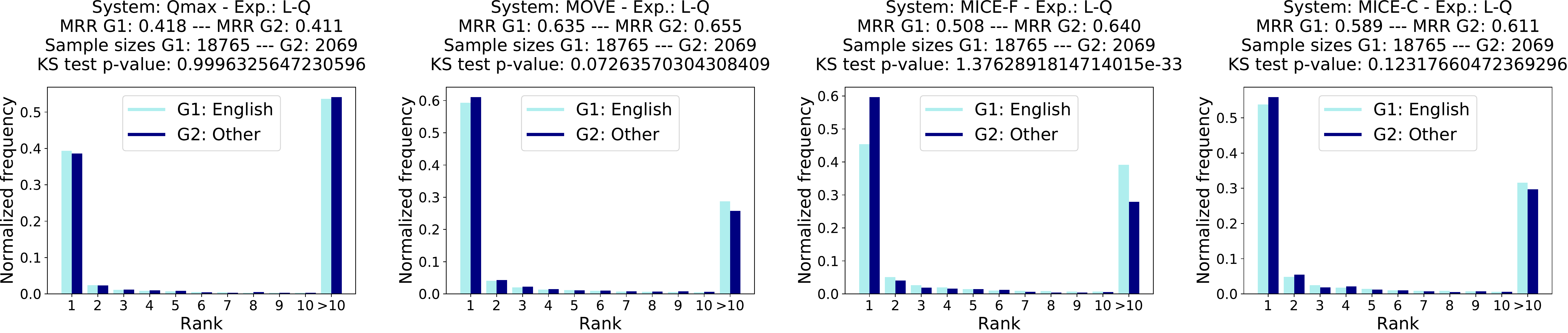}
\vspace{-7mm}
\end{figure*}

\begin{figure*}[h!]
\includegraphics[width=1\linewidth]{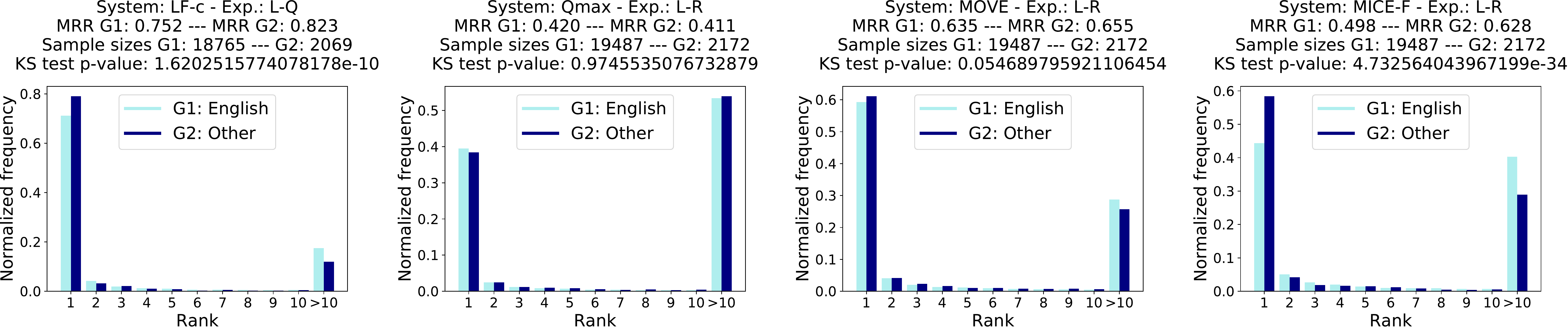}
\vspace{-7mm}
\end{figure*}

\begin{figure*}[h!]
\includegraphics[width=1\linewidth]{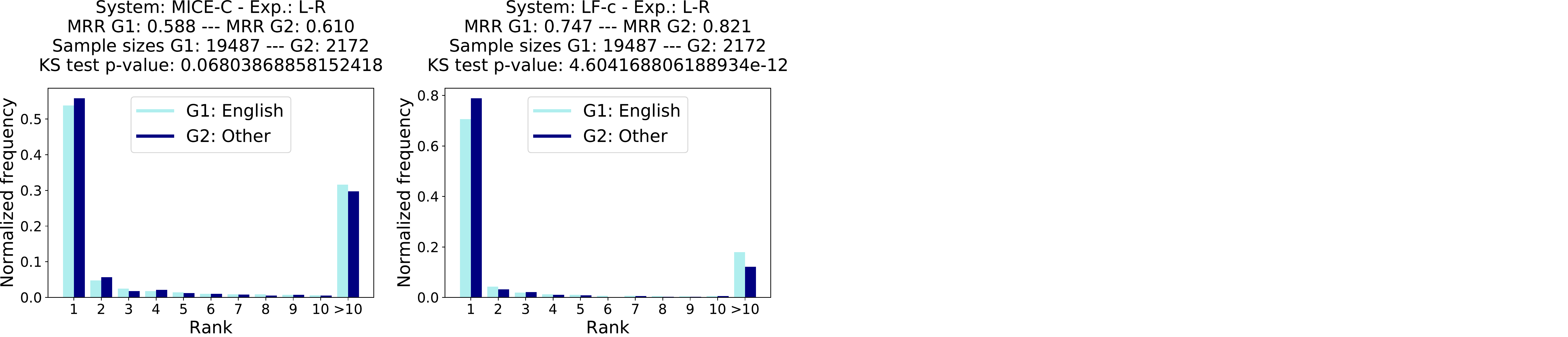}
\vspace{-7mm}
\end{figure*}

\newpage
\subsection{Year experiments}

\begin{figure*}[h!]
\includegraphics[width=1\linewidth]{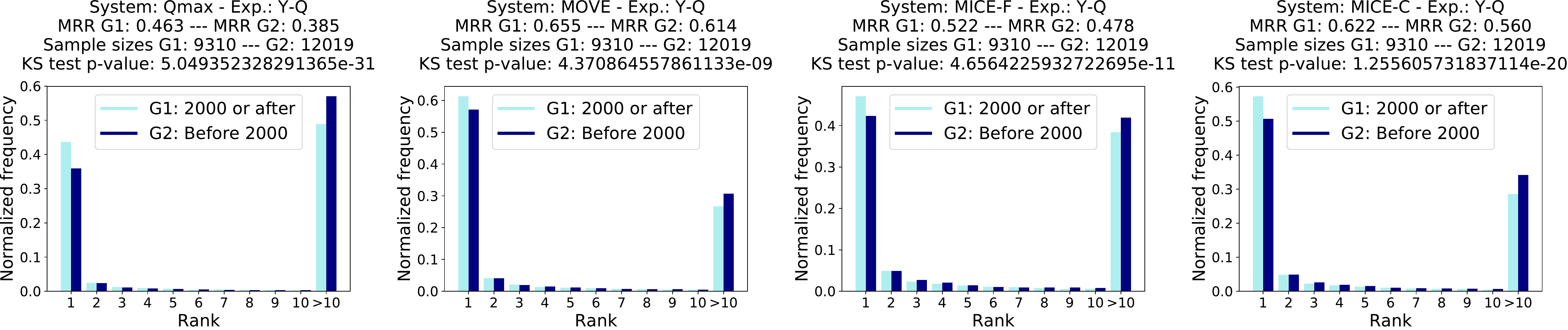}
\vspace{-7mm}
\end{figure*}

\begin{figure*}[h!]
\includegraphics[width=1\linewidth]{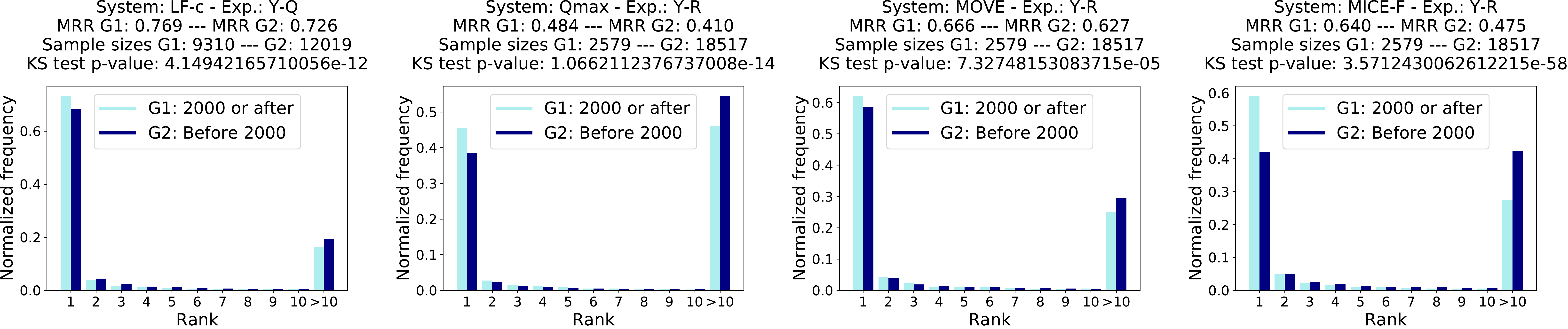}
\vspace{-7mm}
\end{figure*}

\begin{figure*}[h!]
\includegraphics[width=1\linewidth]{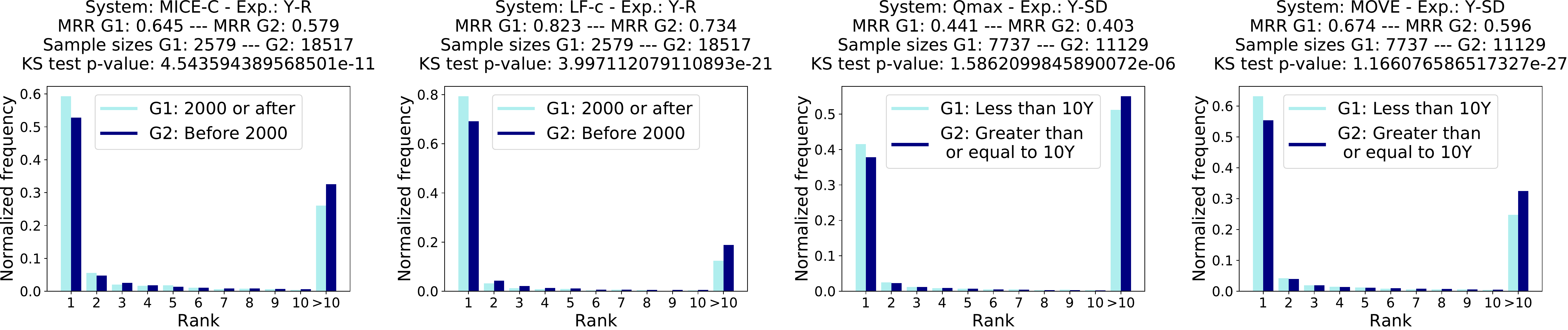}
\vspace{-7mm}
\end{figure*}

\begin{figure*}[h!]
\includegraphics[width=1\linewidth]{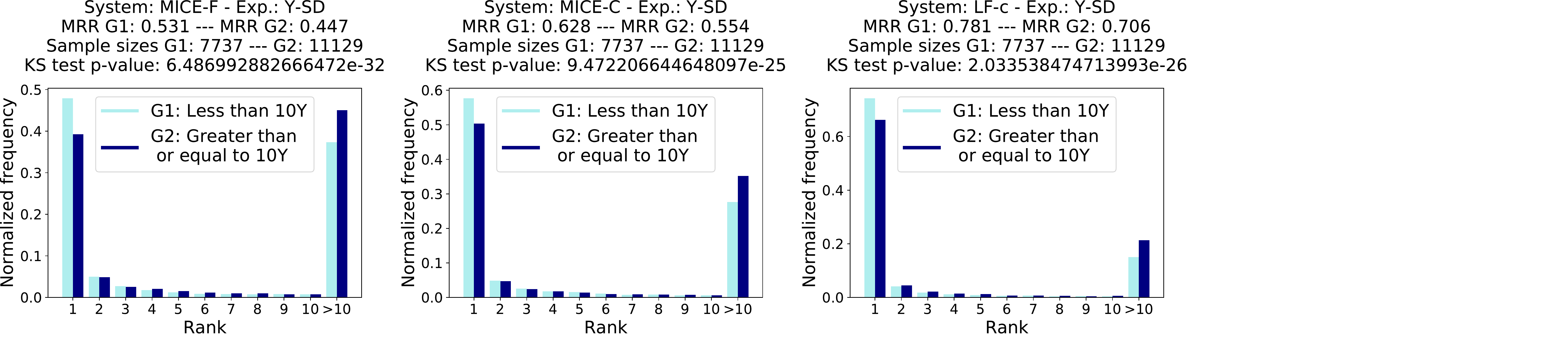}
\vspace{-7mm}
\end{figure*}

\newpage
\subsection{Prevalence experiments}

\begin{figure*}[h!]
\includegraphics[width=1\linewidth]{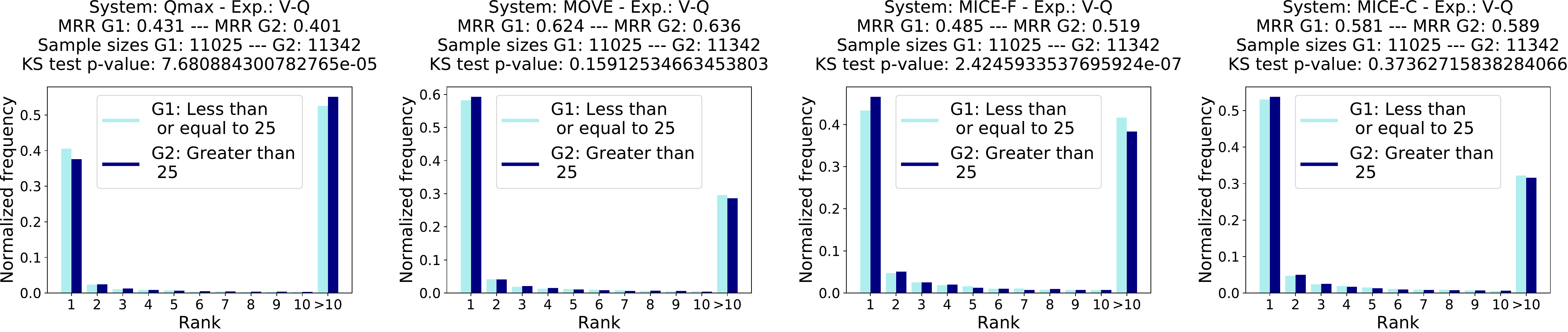}
\vspace{-7mm}
\end{figure*}

\begin{figure*}[h!]
\includegraphics[width=1\linewidth]{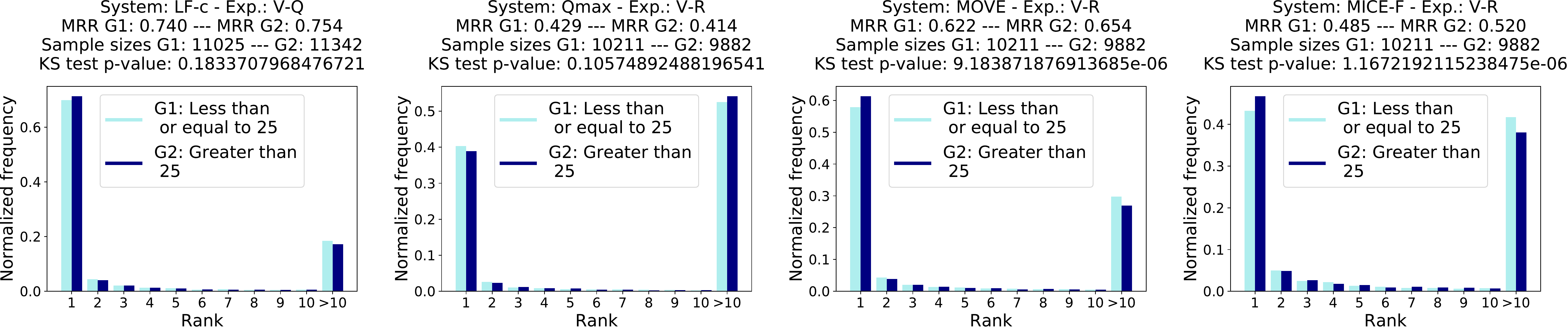}
\vspace{-7mm}
\end{figure*}

\begin{figure*}[h!]
\includegraphics[width=1\linewidth]{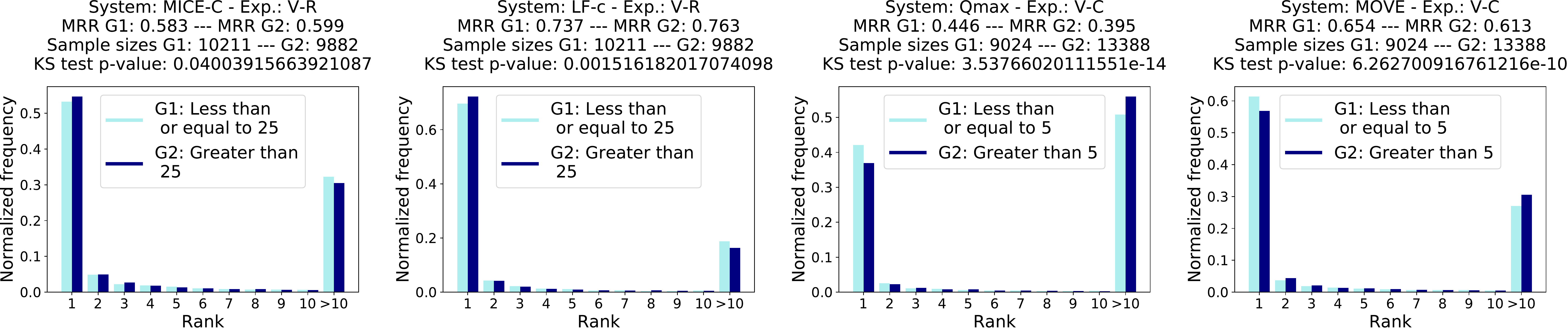}
\vspace{-7mm}
\end{figure*}

\begin{figure*}[h!]
\includegraphics[width=1\linewidth]{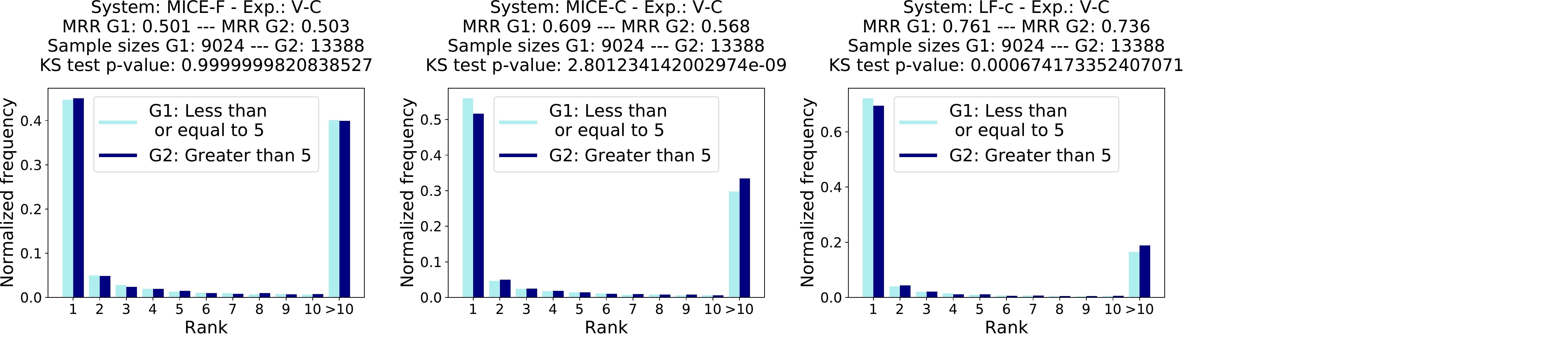}
\vspace{-7mm}
\end{figure*}